\documentclass[10pt]{IEEEtran}
\IEEEoverridecommandlockouts

\usepackage{amsmath,amssymb,amsfonts}
\allowdisplaybreaks
\usepackage{mathtools}
\usepackage{bm}
\usepackage{graphicx}
\usepackage{textcomp}
\usepackage{xcolor}
\usepackage{booktabs}
\usepackage{multirow}
\usepackage{microtype}
\usepackage{nicefrac}
\usepackage{algorithm}
\usepackage{algpseudocode}
\usepackage{tikz}
\usepackage{orcidlink}
\usepackage{comment}
\usepackage{cite}
\usepackage{url}
\usepackage[skip=0pt,font=small]{caption}
\usepackage{physics}

\usepackage{hyperref}

\def\BibTeX{{\rm B\kern-.05em{\sc i\kern-.025em b}\kern-.08em
T\kern-.1667em\lower.7ex\hbox{E}\kern-.125emX}}

\usepackage{setspace}
\begin{document}
\pagestyle{empty}
\title{QuantumChain: Blockchain-Backed Quantum Federated Learning for Financial Fraud Detection}

\author{\IEEEauthorblockN{Epameinondas Douros\textsuperscript{1}, Konstantinos Dalampekis\textsuperscript{1}, Nouhaila Innan\textsuperscript{2,3}, Ioannis Theodonis\textsuperscript{1}, Muhammad Shafique\textsuperscript{2,3}
\IEEEauthorblockA{
\textsuperscript{1}National Technical University of Athens, Athens, Greece\\
\textsuperscript{2}eBRAIN Lab, Division of Engineering, New York University Abu Dhabi (NYUAD), Abu Dhabi, UAE\\
\textsuperscript{3}Center for Quantum and Topological Systems (CQTS), NYUAD Research Institute, NYUAD, Abu Dhabi, UAE\\
Emails: dourosepameinondas@gmail.com, konstantinosdalampekis@gmail.com, nouhaila.innan@nyu.edu, ytheod@mail.ntua.gr, muhammad.shafique@nyu.edu
}}}

\maketitle
\thispagestyle{empty}
\title{QuantumChain: Blockchain-Backed Quantum Federated Learning for Financial Fraud Detection}

\begin{abstract}
Financial fraud detection is challenged by decentralized data, severe class imbalance, and privacy constraints. This paper presents \textit{QuantumChain}, a secure Quantum Federated Learning (QFL) framework that combines hybrid quantum--classical neural networks, encrypted federated aggregation, blockchain-based auditability, and quantum-secure communication. Each client trains a local hybrid model in which a variational quantum circuit is embedded between classical neural layers, while model updates are protected through homomorphic encryption, threshold secret sharing, and QKD-based keying. A permissioned blockchain records aggregation events and supports reputation-weighted trust among participants.
We evaluate \textit{QuantumChain} on financial transaction data using a compact, size-matched classical baseline to isolate the effect of the quantum layer. Results show that the HQNN achieves comparable accuracy while improving fraud-class recall in most settings, reaching $94.6\%$ recall compared with $93.2\%$ for the classical model. The Deep QLayer improves performance in full-data settings, suggesting that added circuit depth helps recover representational capacity when the shallow circuit becomes limited. Mixed-state simulations further show that the recall trend persists under non-ideal quantum evolution. In federated deployment with $10$ heterogeneous clients, global accuracy increases from $97.7\%$ to $98.8\%$ over five rounds before stabilizing. These results show that \textit{QuantumChain} can integrate depth-aware hybrid quantum models into a secure federated fraud-detection pipeline while maintaining stable global convergence.
\end{abstract}


\section{Introduction}

Financial fraud remains a major challenge for modern banking systems as digital payments, instant transfers, and cross-platform financial services continue to expand. Recent reports indicate that criminals stole £1.28 billion through payment fraud in the UK in 2025, a 4\% increase over 2024~\cite{ukfinanceAnnualFraud}, while global payment card fraud losses reached approximately \$33 billion~\cite{globenewswireGlobalCard}. These losses highlight the need for fraud detection systems that can identify rare fraudulent transactions across institutions without exposing sensitive customer data. Although machine learning has improved fraud detection, many models still rely on centralized data collection and struggle with severe class imbalance, where recall for the minority fraud class is more important than small gains in accuracy.

Federated Learning (FL) enables institutions to train shared models while keeping raw data local~\cite{uddin2025systematic}. However, standard FL does not fully address financial security and trust requirements. Model updates may leak private information, unreliable clients can degrade aggregation quality, and aggregation decisions are often difficult to audit. These limitations are critical in banking environments, where confidentiality, integrity, accountability, and trust must be preserved during training.

Quantum Machine Learning (QML) offers a promising direction for richer feature representation under such constraints~\cite{rodriguez2025survey}. Hybrid quantum--classical neural networks (HQNNs) use variational quantum circuits as trainable feature maps and can improve minority-class sensitivity in imbalanced transaction data. Existing studies mainly examine quantum models as standalone classifiers or as simple federated clients~\cite{innan2024financial,innan2024financial1,alami2025fid,el2026comparative}. Prior QFL fraud-detection work has focused on combining quantum neural models with federated training~\cite{innan2025qfnn,sawaika2025privacy,abbassi2025adaptive}, but secure aggregation, auditable coordination, QKD-secured communication, and trust-aware client weighting remain insufficiently explored within a single framework.

\begin{figure*}[t]
    \centering
    \includegraphics[width=1\linewidth]{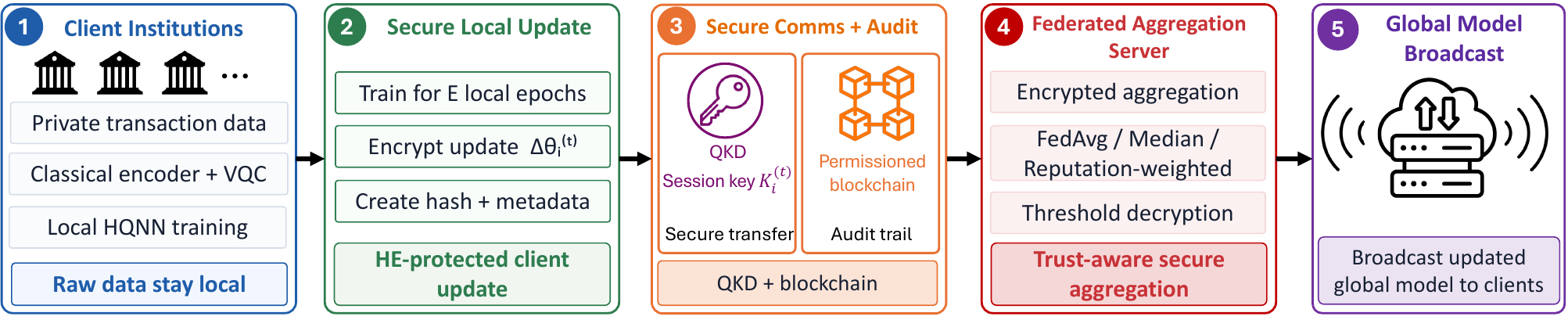}
\caption{Overview of \textit{QuantumChain}, including local HQNN training, encrypted update submission, blockchain-based auditability, QKD-derived session keys, and secure federated aggregation.}
    \label{fig:method}
\end{figure*}

To address this gap, we propose \textit{QuantumChain}, a blockchain-backed quantum federated learning (QFL) framework for secure financial fraud detection. As shown in Fig.~\ref{fig:method}, each institution trains a local HQNN on private transaction data, where a variational quantum layer is embedded inside a compact classical neural model. Model updates are protected through homomorphic encryption and threshold secret sharing, while a permissioned blockchain records update commitments and aggregation metadata. QKD-derived session keys secure communication across training rounds, and reputation-weighted aggregation reduces the effect of unreliable clients.

The main contributions are:
\begin{enumerate}
\item We propose \textit{QuantumChain}, a secure QFL framework that combines HQNN-based fraud detection, encrypted aggregation, blockchain auditability, QKD-secured communication, and reputation-weighted trust.
\item We design and evaluate Shallow and Deep QLayer variants that share the same encoding and measurement readout but differ in circuit depth, enabling a controlled study of quantum-layer expressivity for fraud recall.
\item We evaluate the framework on financial transaction data, showing that HQNNs match a compact classical baseline in accuracy while improving fraud-class recall in most settings, and that the global federated model converges stably across communication rounds.
\end{enumerate}

The HQNN improves average fraud recall while maintaining comparable accuracy, with the Deep QLayer reaching $94.6\%$ recall and the federated model reaching $98.8\%$ global accuracy.
\section{Methodology}
\label{sec:methodology}

\textit{QuantumChain} targets collaborative fraud detection across $K$ clients, where each client represents a financial institution that keeps transaction data local, as illustrated in Fig.~\ref{fig:method}.
\subsection{Problem Formulation}

Let $\mathcal{C}=\{1,\ldots,K\}$ denote the set of clients. Client $i$ owns a private dataset
$\mathcal{D}_i=\{(x_j^{(i)},y_j^{(i)})\}_{j=1}^{n_i}$, where $x_j^{(i)}\in\mathbb{R}^{d}$ is a transaction feature vector and $y_j^{(i)}\in\{0,1\}$ denotes a legitimate or fraudulent transaction. The goal is to learn a global model $f_\theta$ without sharing raw data:
\begin{equation}
\begin{aligned}
\min_{\theta}\mathcal{L}(\theta)
= \sum_{i=1}^{K}\frac{n_i}{N}\mathcal{L}_i(\theta), \quad
\mathcal{L}_i(\theta)
= \frac{1}{n_i}\sum_{(x,y)\in\mathcal{D}_i}
\ell(f_\theta(x),y),
\end{aligned}
\end{equation}
where $\theta$ denotes the global model parameters, $N=\sum_i n_i$ is the total number of samples, and $\ell(\cdot)$ is the binary cross-entropy loss.

\subsection{Hybrid Quantum--Classical Local Model}

\begin{figure*}[t]
    \centering
    \includegraphics[width=1\linewidth]{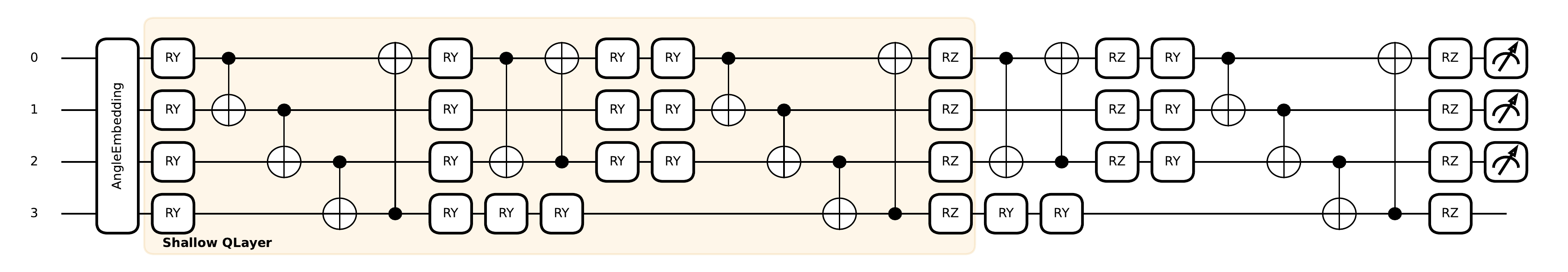}
\caption{Shallow and Deep QLayer architectures. The highlighted region denotes the Shallow QLayer embedded within the Deep QLayer, with shared AngleEmbedding input encoding and Pauli-$Z$ readout.}
    \label{fig:qlayer_deep_shallow}
\end{figure*}

Each client trains an HQNN composed of a classical encoder, a variational quantum layer, and a classical output layer. The encoder maps $x\in\mathbb{R}^{d}$ to $q$ rotation angles:
$\alpha(x)=Wx+b,$
where $W\in\mathbb{R}^{q\times d}$ and $b\in\mathbb{R}^{q}$ are trainable encoder parameters. The resulting angle vector,
$\alpha(x)=[\alpha_1(x),\ldots,\alpha_q(x)],$
is used for angle encoding as follows:
$|\psi_{\mathrm{enc}}(x)\rangle=
\bigotimes_{k=1}^{q}R_Y!\left(\alpha_k(x)\right)|0\rangle.$
A variational circuit $U(\varphi)$ with trainable quantum parameters $\varphi$ produces
$|\psi_{\varphi}(x)\rangle = U(\varphi)|\psi_{\mathrm{enc}}(x)\rangle.$
As shown in Fig.~\ref{fig:qlayer_deep_shallow}, we use two circuit variants: a Shallow QLayer and a Deep QLayer. Both use the same qubit count, input encoding, and Pauli-$Z$ measurement scheme. The Deep QLayer contains the shallow circuit as a substructure and appends extra trainable rotation and entangling blocks, increasing feature extraction capacity at the cost of additional parameters and quantum operations.

The measured quantum features are
$m_{\varphi}(x)=\left[\langle Z_1\rangle,\langle Z_2\rangle,\dots,\langle Z_q\rangle\right]$,
where $\langle Z_k\rangle$ is the Pauli-$Z$ expectation value on qubit $k$. These features are passed to a classical classifier:
$f_{\theta}(x)=\sigma\!\left(a^\top m_{\varphi}(x)+c\right)$,
where $a$ and $c$ are trainable classical parameters and $\sigma(\cdot)$ is the sigmoid activation. Gradients through the variational circuit are computed using the parameter-shift rule.

\subsection{Secure Federated Optimization}

Training proceeds over communication rounds $t=0,1,\ldots$. At each round, the server sends the current global model $\theta^{(t)}$ to selected clients $\mathcal{S}_t$. Each client performs $E$ local epochs: $\theta_i^{(t,e+1)}
=
\theta_i^{(t,e)}
-
\eta\nabla_\theta \mathcal{L}_i(\theta_i^{(t,e)}),$ for $ e=0,\ldots,E-1,$
where $\eta$ is the learning rate. The local update is
$\Delta\theta_i^{(t)}=\theta_i^{(t,E)}-\theta^{(t)}$.
To protect update confidentiality, client $i$ encrypts its update as
$c_i^{(t)}=\mathrm{Enc}_{\mathrm{pk}}(\Delta\theta_i^{(t)})$,
where $\mathrm{pk}$ is the public encryption key. Encrypted aggregation is performed as $c_{\mathrm{agg}}^{(t)}
=
\bigoplus_{i\in\mathcal{S}_t}
w_i^{(t)}\odot c_i^{(t)},$ and $\sum_{i\in\mathcal{S}_t}w_i^{(t)}=1,
$
where $w_i^{(t)}$ is the aggregation weight, $\oplus$ denotes homomorphic addition, and $\odot$ denotes scalar multiplication in the encrypted domain. The aggregate is decrypted through threshold secret sharing, so no single party holds the full decryption key.

We study FedAvg, componentwise median, and reputation-weighted aggregation. FedAvg weights clients by data size, while the median rule reduces the effect of extreme updates. For reputation-weighted aggregation, client $i$ receives a trust score $\kappa_i^{(t)}$ based on update quality, validation behavior, or update consistency:
\begin{equation}
w_i^{(t)}=
\frac{\kappa_i^{(t)}\mathbf{1}\{\kappa_i^{(t)}\geq \tau_t\}}
{\sum_{j\in\mathcal{S}_t}\kappa_j^{(t)}
\mathbf{1}\{\kappa_j^{(t)}\geq \tau_t\}},
\end{equation}
where $\tau_t$ is the round-dependent trust threshold and $\mathbf{1}\{\cdot\}$ is an indicator function. This reduces the influence of unreliable updates while retaining useful client contributions.

\subsection{Blockchain and QKD Security Layer}

For auditability, each encrypted update is committed to a permissioned blockchain:
$h_i^{(t)}=H\!\left(c_i^{(t)}\Vert \mathrm{meta}_i^{(t)}\right)$,
where $H(\cdot)$ is a hash function, $\Vert$ denotes concatenation, and $\mathrm{meta}_i^{(t)}$ contains the round index, client identity, timestamp, and update identifier. The blockchain stores only hashes and non-sensitive metadata; plaintext updates and raw data are never stored on-chain. In our implementation, this layer is simulated using a local Ethereum/Ganache environment.

For communication security, client $i$ obtains a fresh round-specific session key:
$K_i^{(t)}\leftarrow \mathrm{QKD}(i,\mathrm{server})$.
We model QKD as a key-service abstraction rather than a deployed optical QKD network. The assumed network contains an authenticated classical channel and either point-to-point QKD links or trusted QKD nodes. Since QKD does not authenticate the classical channel by itself, client identities and QKD messages must be authenticated using pre-shared credentials, digital signatures, or post-quantum authentication.

Each QKD session estimates the quantum bit error rate $Q_i^{(t)}$. A key is accepted only if $Q_i^{(t)}\leq Q_{\max}$, where $Q_{\max}$ is the protocol threshold. Let $B_i^{(t)}$ be the number of bits required to secure client $i$'s transmission, $R_i$ the secret-key rate, and $T_i^{(t)}$ the available key-generation time. The session is feasible when $R_iT_i^{(t)}\geq B_i^{(t)}.$
In \textit{QuantumChain}, QKD-derived keys protect transport channels, while homomorphic encryption protects update contents during aggregation. Compared with post-quantum key exchange, QKD provides eavesdropping detection through $Q_i^{(t)}$, but requires QKD infrastructure; post-quantum key exchange can therefore serve as a fallback.

\subsection{Training Procedure and Security Properties}

Algorithm~\ref{alg:qchain} summarizes one communication round of \textit{QuantumChain}. The cryptographic layer is model-agnostic, so the same secure pipeline can support both classical and hybrid quantum--classical local models.

\begin{algorithm}[t]
\caption{One Round of \textit{QuantumChain}}
\label{alg:qchain}
\footnotesize
\begin{algorithmic}[1]
\State Server broadcasts $\theta^{(t)}$ to selected clients $\mathcal{S}_t$.
\For{each client $i\in\mathcal{S}_t$}
\State Establish $K_i^{(t)}\leftarrow\mathrm{QKD}(i,\mathrm{server})$.
\State Train local HQNN for $E$ epochs on $\mathcal{D}_i$.
\State Compute $\Delta\theta_i^{(t)}=\theta_i^{(t,E)}-\theta^{(t)}$.
\State Encrypt update $c_i^{(t)}=\mathrm{Enc}_{\mathrm{pk}}(\Delta\theta_i^{(t)})$.
\State Commit $h_i^{(t)}=H(c_i^{(t)}\Vert\mathrm{meta}_i^{(t)})$ on-chain.
\State Send $c_i^{(t)}$ using the QKD-derived session key.
\EndFor
\State Server aggregates encrypted updates to obtain $c_{\mathrm{agg}}^{(t)}$.
\State Committee performs threshold decryption.
\State Server updates the global model and records the aggregate commitment.
\end{algorithmic}
\end{algorithm}
The design provides layered protection. FL keeps raw transaction data local; homomorphic encryption protects individual updates during aggregation; threshold secret sharing prevents any single party from decrypting the aggregate alone; blockchain commitments provide a tamper-evident audit trail; and QKD-derived session keys protect communication in each round. This design does not assume that any single mechanism solves all FL attacks; instead, each component protects a specific part of the training pipeline.
\section{Experimental Setup and Results}
\label{sec:results}
The empirical evaluation studies local HQNN performance, the effect of quantum-circuit depth, and federated integration under the encrypted \textit{QuantumChain} pipeline. The goal is not to claim state-of-the-art fraud detection against large models such as boosted trees, deep ensembles, or graph-based detectors. Instead, we use a compact matched neural baseline to isolate the effect of replacing the intermediate classical representation with the variational quantum layer described in Sec.~\ref{sec:methodology}. The experiments use a publicly available credit-card fraud dataset\footnote{\url{https://www.kaggle.com/datasets/ealaxi/paysim1}}.

We compare the classical baseline against two HQNN variants, Shallow QLayer and Deep QLayer, defined in Fig.~\ref{fig:qlayer_deep_shallow}. The two HQNN variants share the same input encoding and measurement readout but differ in circuit depth. All models are implemented using PennyLane and TensorFlow/Keras. Ideal simulations use \texttt{default.qubit}, while mixed-state robustness is evaluated with \texttt{default.mixed}. We vary the dataset fraction $\mathrm{DS}$, batch size $\mathrm{BS}$, and number of local training epochs $E$, with $\mathrm{DS}\in\{30\%,100\%\}$, $\mathrm{BS}\in\{128,256,512\}$, and $E\in\{3,5,10\}$. Each local configuration is averaged over five independent runs. In Tables~\ref{tab:local_results} and~\ref{tab:global_results}, Acc., Rec., and F1 denote accuracy, fraud-class recall, and fraud-class F1-score, respectively. Since fraud detection is highly imbalanced, Rec. is used as the primary metric.

\subsection{Client-Level Model Performance}

Table~\ref{tab:local_results} reports client-level results before federated aggregation. These experiments provide a controlled comparison between the compact classical model and the HQNN under matched training conditions. Settings S1--S3 evaluate the Shallow QLayer, while S4--S6 evaluate the Deep QLayer. The Shallow QLayer improves Rec. in reduced-data settings, but slightly underperforms the classical baseline in S3, suggesting limited capacity under larger data complexity. The Deep QLayer improves Rec. in S4--S6, reaching $94.60\%$ in S6 compared with $93.20\%$ for the classical model, although with lower F1, indicating a sensitivity--precision trade-off.
\begin{table}[t]
\centering
\caption{Client-level results on \texttt{default.qubit}. $\mathrm{DS}$ denotes dataset fraction, $\mathrm{BS}$ denotes batch size, and $E$ denotes local training epochs. Values are averaged over five runs.}
\label{tab:local_results}
\footnotesize
\setlength{\tabcolsep}{3.2pt}
\renewcommand{\arraystretch}{1.08}
\begin{tabular}{c c c l c c c}
\toprule
\textbf{Set.} & \textbf{DS} & \textbf{BS,E} & \textbf{Model} & \textbf{Acc.} & \textbf{Rec.} & \textbf{F1} \\
\midrule
S1 & 30\% & 128,5 & Classical & \textbf{98.80} & 91.20 & \textbf{92.60} \\
   &      &       & HQNN-shallow & 98.40 & \textbf{92.00} & 91.20 \\
S2 & 30\% & 256,5 & Classical & 98.30 & 86.80 & 90.50 \\
   &      &       & HQNN-shallow & \textbf{98.90} & \textbf{90.90} & \textbf{92.70} \\
S3 & 100\% & 512,3 & Classical & \textbf{98.70} & \textbf{90.26} & \textbf{92.80} \\
   &       &       & HQNN-shallow & 98.68 & 88.88 & 92.38 \\
S4 & 100\% & 256,5 & Classical & 98.36 & 87.32 & 90.88 \\
   &       &       & HQNN-deep & \textbf{98.96} & \textbf{90.24} & \textbf{92.72} \\
S5 & 100\% & 512,3 & Classical & 98.00 & 82.80 & 88.20 \\
   &       &       & HQNN-deep & \textbf{98.20} & \textbf{86.60} & \textbf{90.40} \\
S6 & 100\% & 256,10 & Classical & \textbf{99.00} & 93.20 & \textbf{95.20} \\
   &       &        & HQNN-deep & 98.40 & \textbf{94.60} & 93.20 \\
\bottomrule
\end{tabular}
\end{table}

Across the six representative settings, the classical baseline obtains an average Acc. of $98.53\%$, Rec. of $88.60\%$, and F1 of $91.70\%$. The HQNN obtains an average Acc. of $98.59\%$, Rec. of $90.54\%$, and F1 of $92.10\%$. The main improvement is therefore observed in fraud-class Rec., which is the primary metric for this imbalanced task.

\subsection{Mixed-State Robustness}

To test whether the Rec. trend persists under non-ideal quantum evolution, we repeat the comparison using \texttt{default.mixed}. Due to the higher computational cost of mixed-state simulation, we use a constrained setting with $\mathrm{DS}=1\%$, $\mathrm{BS}=64$, $E=5$, and the Deep QLayer. The classical model achieves $98.20\%$ Acc., $81.53\%$ Rec., and $88.53\%$ F1. Under the same setting, the HQNN achieves $98.00\%$ Acc., $83.33\%$ Rec., and $89.33\%$ F1. Although absolute Rec. is lower than in the ideal full-data setting, the HQNN keeps a Rec. advantage, suggesting that the quantum layer remains useful under mixed-state simulation.

\subsection{Federated Global Model Performance}

After the client-level study, we deploy the HQNN inside the federated \textit{QuantumChain} pipeline with $10$ heterogeneous clients. Each client trains locally for $3$ epochs per communication round and sends encrypted model updates to the aggregator; raw transaction data remain local. The server updates a global model using the selected aggregation rule, and the aggregated model is evaluated on the combined test set after each global round.
As presented in Table~\ref{tab:global_results}, the global model shows stable convergence over five communication rounds. Global Acc. increases from $97.73\%$ in round~1 to $98.80\%$ in round~4, and remains stable at $98.73\%$ in round~5. The logged workflow confirms that each round performs encrypted update submission, blockchain commitment, encrypted aggregation, QKD-keyed model redistribution, and global evaluation. This shows that the HQNN local models can be integrated into the encrypted federated pipeline without destabilizing the global model.

\begin{table}[t]
\centering
\caption{Global-model accuracy during federated training.}
\label{tab:global_results}
\footnotesize
\setlength{\tabcolsep}{8pt}
\renewcommand{\arraystretch}{1.08}
\begin{tabular}{c c c c c}
\toprule
\textbf{Round 1} & \textbf{Round 2} & \textbf{Round 3} & \textbf{Round 4} & \textbf{Round 5} \\
\midrule
97.73\% & 98.56\% & 98.66\% & \textbf{98.80\%} & 98.73\% \\
\bottomrule
\end{tabular}
\end{table}

At the final round, the client-level confusion matrices yield an average Acc. of $98.73\%$, fraud-class precision of $98.11\%$, Rec. of $97.44\%$, and F1 of $97.77\%$ across the $10$ clients, indicating stable post-aggregation behavior.

\subsection{Aggregation Strategy Analysis}

We compare FedAvg, componentwise median, and reputation-weighted aggregation. FedAvg serves as the standard averaging baseline, while median aggregation reduces the effect of extreme client updates. In the reputation-weighted setting, client trust scores are used to adjust aggregation weights and reduce the influence of unstable or low-quality updates. In our experiments, this strategy provided the best observed balance between convergence stability and robustness, suggesting that trust-aware aggregation can be useful when hybrid quantum clients produce updates with different reliability.

\subsection{Discussion}

The results support three findings. First, the HQNN broadly matches the compact classical baseline in Acc. while improving Rec. in most client-level settings. This is important for fraud detection because missing fraudulent samples is more costly than small changes in Acc. Second, circuit depth matters: the Shallow QLayer helps in reduced-data settings, while the Deep QLayer is more effective on the full dataset. Third, the federated results show that the hybrid local model can be integrated into the secure \textit{QuantumChain} pipeline while preserving stable global convergence.
These findings should be interpreted as a controlled hybrid-model study, not as a claim of superiority over stronger fraud-detection architectures. The novelty lies in combining a depth-aware HQNN design with encrypted federated training and trust-aware aggregation, while showing that the quantum layer can improve fraud-class sensitivity under matched local-model conditions.
\section{Conclusion}

This paper introduced \textit{QuantumChain}, a secure quantum-enhanced federated learning framework for financial fraud detection. Under a compact matched-baseline comparison, the HQNN improves average fraud-class recall from $88.60\%$ to $90.54\%$ while maintaining comparable accuracy. The federated model remains stable across communication rounds, reaching $98.80\%$ global accuracy. Future work will test the framework on quantum hardware, larger heterogeneous datasets, and stronger adversarial settings.

\setstretch{0.90}{
\section*{Data and Code Availability} 
\small The code developed for this paper is publicly available at \url{https://github.com/EpameinondasDouros/QFL-Fraud-Detection}.
\section*{Acknowledgment}
\small  This work was supported in part by the NYUAD Center for Quantum and Topological Systems (CQTS), funded by Tamkeen under the NYUAD Research Institute grant CG008,  and by the NYUAD Center for CyberSecurity (CCS), funded by Tamkeen under the NYUAD Research Institute Award G1104.
\bibliographystyle{IEEEtran}

\bibliography{refs}}

\end{document}